\pdfoutput=1
\documentclass[journal,twocolumn]{IEEEtran}
\usepackage[utf8]{inputenc}
\usepackage{amsmath,amsfonts,amssymb,amsthm}
\usepackage{cite}
\usepackage{graphicx}
\usepackage{hyperref}
\usepackage{xcolor}
\usepackage[small]{caption}
\usepackage{booktabs}
\usepackage{balance}
\usepackage{subfig}

\newcommand{\tr}{\mathrm{Tr}} 
\newcommand{\sinc}{\operatorname{sinc}} 
\newcommand{\E}{\mathbb{E}} 
\newcommand{\V}{\mathrm{Var}} 

\title{Ambiguity Function Analysis of Affine Frequency Division Multiplexing for Integrated Sensing and Communication}
\author{Ebrahim Bedeer
	\thanks{E. Bedeer is with the Department of Electrical and Computer Engineering, University of Saskatchewan, Saskatoon, Canada S7N5A9. Email: e.bedeer@usask.ca.}
}

\begin{document}
	\maketitle

\begin{abstract}
	Affine frequency division multiplexing (AFDM) is a chirp-based multicarrier waveform that was recently proposed for communication over doubly dispersive channels. Given its chirp nature, AFDM is expected to have superior sensing capabilities compared to orthogonal frequency division multiplexing (OFDM) and is thus a promising candidate for integrated sensing and communication (ISAC) applications. In this paper, we derive a closed-form expression for the ambiguity function of AFDM waveforms modulated with $M$-ary quadrature amplitude modulation (QAM) data symbols. 
	We determine the condition on the chirp rate of the AFDM waveform that minimizes the sidelobes in the delay/range domain in the presence of random $M$-ary QAM symbols, thereby improving overall sensing performance. 
	Additionally, we find an approximate statistical distribution for the magnitude of the derived ambiguity function. Simulation results are presented to evaluate the sensing performance of the AFDM waveform for various system parameters and to compare its peak-to-sidelobe ratio (PSLR) and integrated sidelobe ratio (ISLR) with those of OFDM.
	
\end{abstract}

\begin{IEEEkeywords}
	Affine frequency division multiplexing (AFDM), ambiguity function, integrated sensing and communication (ISAC)
\end{IEEEkeywords}

\section{Introduction}\label{sec:Introduction}

Integrated sensing and communication (ISAC) is envisioned to be one of the key technologies for 6G \cite{gonzalez2024integrated}. ISAC systems enable spectral and hardware coexistence between sensing and communication devices. Such coexistence improves the spectral efficiency and reduce cost and size for future wireless networks. The integration of sensing and communication is natural given that many communication devices connected to cellular wireless networks are equipped with sensors and already perform sensing tasks, e.g., connected vehicles. Additionally, communication devices are expected to operate at higher frequencies and bandwidths and will be equipped with large antenna arrays which result in signal processing at the transceiver that facilitates sensing. While several challenges exist in designing efficient ISAC systems,  the design of waveforms that enable both sensing and communication has gained considerable attention  \cite{zhou2022integrated}.

Orthogonal frequency division multiplexing (OFDM) is currently used in several wireless systems, e.g., 4G LTE/LTE-Advanced and 5G NR, due to numerous advantages.
Additionally, OFDM has been extensively studied for radar and sensing applications \cite{zhou2022integrated}. However, OFDM has limitations that include its poor performance over doubly dispersive channels where the orthogonality between its subcarriers is lost, and hence, results in inter-carrier interference (ICI), due to the high Doppler frequency shifts. Accordingly, there is a growing interest in the research community in finding alternative waveforms to OFDM that maintain its benefits while avoiding some of its drawbacks.

Affine Frequency Division Multiplexing (AFDM) was recently introduced to address the poor performance of OFDM over doubly dispersive channels \cite{bemani2023affine}. AFDM is based on multicarrier chirp signals where the modulation symbols are spread over the entire time-frequency plane. Hence, it is robust to doubly dispersive channel conditions, as can be seen from its achieved optimal diversity order. Since chirp signals are inherently suitable for radar and sensing applications,
AFDM is expected to be one of the candidate waveforms for ISAC.

Recent studies, e.g., \cite{ni2022afdm, bemani2024integrated, zhu2024afdm}, investigated the applications of AFDM in ISAC. In particular, the authors in \cite{ni2022afdm} proposed two techniques to estimate the range and velocity of targets using the entire AFDM frame with modulated random information symbols. Unlike \cite{ni2022afdm}, the authors in \cite{bemani2024integrated} showed that a single pilot symbol and its guard interval in the discrete affine domain can identify all the delay and Doppler components of the channel. In \cite{zhu2024afdm}, the authors analyzed the zero-Doppler response of the ambiguity function and concluded that to achieve  good sensing performance, the chirp rate should be selected as in \cite{bemani2023affine}. The analysis of the zero-Doppler response of the ambiguity function  in \cite{zhu2024afdm} did not consider that the AFDM waveform  carries modulation symbols; rather, it is assumed that the symbols  have a fixed value of 1.  We will show in this paper that the selection of the chirp rate as in \cite{zhu2024afdm, bemani2023affine} is not optimal to reduce the sidelobes in the delay/range domain in the presence of random $M$-ary QAM symbols, and we will derive the optimal chirp rate of the AFDM waveform for sensing applications with random data symbols.
The authors in \cite{berggren2022joint} studied the ambiguity function of a chirp-based multicarrier waveform that carries a pre-designed transmit sequence, i.e., not random modulation data symbols, to reduce the peak-to-average-power-ratio (PAPR) of the waveform. Additionally, the authors in \cite{berggren2022joint} studied the effect of $M$-ary phase shift keying (PSK) modulated data symbols on the behavior of the ambiguity function.

In this paper, we  derive a closed-form expression for the ambiguity function  of AFDM waveforms modulated with $M$-ary quadrature amplitude modulation (QAM) data symbols to characterize the overall radar accuracy. We determine the optimal AFDM waveform chirp rate that minimizes the sidelobes in the delay/range domain, and hence, improve the overall AFDM sensing performance. We additionally show that the magnitude of the derived ambiguity function is well approximated by the Rice distribution,  and we identify the distribution parameters. Simulation results are presented to evaluate the radar global performance of the AFDM waveform and to verify the accuracy of the approximate Rice distribution. The results also demonstrate the superiority of AFDM over OFDM in terms of the peak-to-sidelobe ratio (PSLR) and integrated sidelobe ratio (ISLR). Additionally, the results show that, for a fixed total bandwidth, increasing the number of chirps reduces both the PSLR and ISLR; while both the PSLR and ISLR are insensitive to changing the modulation order.

The remainder of this paper is organized as follows. Section~\ref{sec:system_model} describes the system model of AFDM. We derive the ambiguity function of AFDM waveforms in Section \ref{sec:AF}; while its magnitude analysis and approximate distribution derivation are presented in Section~\ref{sec:proposed}. In Section~\ref{sec:PSLR_ISLR}, we define and calculate the PSLR and ISLR. Simulation results are provided in Section \ref{sec:simulation}, and Section~\ref{sec:conclu} concludes the paper.

\emph{Notations:} Throughout the paper, lower-case and upper-case boldface letters denote vectors and matrices, respectively. The notation $\mathbf{A}_{a_1,a_2}$ represents the $(a_1, a_2)$ element of matrix $\mathbf{A}$. {The transpose of  vector $\mathbf{a}$ is denoted as $\mathbf{a}^{\rm{T}}$,} the  Hermitian of  matrix $\mathbf{A}$ is denoted as $\mathbf{A}^{\text{H}}$, and the complex conjugate of $a$ is $a^*$. 
We use $\tr\left({\mathbf{A}}\right)$ and $\operatorname{diag}\left({\mathbf{A}}\right)$ to denote the trace and the diagonal vector of matrix $\mathbf{A}$.
We use $\mathbf{I}_{N}$  to represent the identity matrix of size $N\ \times\ N $, $\mathbb{C}^{N_1 \times N_2}$ to represent the set of complex matrices of size $N_1 \times N_2$, and $\|\mathbf{a}\|_2$ to represent the 2-norm of vector $\mathbf{a}$.  {We denote a complex Gaussian random variable with  mean $\mu$ and variance $\sigma^2$  as $\mathcal{C} \mathcal{N}(\mu,\,\sigma^2)$}. Further, $\E[\cdot]$ and $\V[\cdot]$  represent the expectation and the variance, respectively. The set of all integer is denoted by  $\mathbb{Z}$. We use calligraphic fonts, i.e., $\mathcal{A}$ to denote non-standard sets, and to exclude an element $a$ from the set $\mathcal{A}$, we use $\mathcal{A} \backslash a$.

\section{System Model}\label{sec:system_model}

We consider an AFDM ISAC system that operates at a bandwidth $B = N \Delta_f$, where $N$ and $\Delta_f$ represent the number of chirp subcarriers and the chirp subcarrier spacing, respectively. The useful transmit AFDM symbol duration is $T = 1/\Delta_f$. The random communication symbols $x_n$, $n = 0, \dots, N-1$, in the affine domain are assumed to be drawn from an $M$-ary QAM constellation. We assume that the $M$-ary QAM constellations are normalized to have unit power, i.e., $\E\left(|x_n|^2\right) = 1$, and are proper, i.e., $\E\left(x_n\right) = 0$ and $\E\left(x_n^2\right) = 0$.
Using the inverse discrete affine Fourier transform (IDAFT), the modulated signal in the discrete-time domain is written as \cite{bemani2023affine}
\begin{IEEEeqnarray}{RCL}
	s[n]&=& \frac{1}{\sqrt{N}}\sum_{m = 0}^{N-1} x_m e^{j2\pi\left(c_1 n^2 + c_2 m^2 + nm/N\right)}, \nonumber \\ && \hfill n = 0, ..., N-1,
\end{IEEEeqnarray}	
where $c_1$ and $c_2$ are the discrete affine Fourier transform parameters, and $c_1$ represents the AFDM chirp rate. A chirp-periodic prefix (CPP) is then added to  the discrete-time transmit signal $s[n]$ to mitigate the inter-symbol interference (ISI) due to the multipath propagation channel. The CPP signal is defined as \cite{bemani2023affine}
\begin{IEEEeqnarray}{RCL}
	s[n]&=& s[N + n] e^{-j2\pi c_1 N (N^2 + 2N n)}, \quad n = - L_{\text{cp}}, ..., - 1,	\IEEEeqnarraynumspace
\end{IEEEeqnarray}
where $L_{\text{cp}}$ is an integer value greater than or equal to the sampled maximum delay spread of the channel. The discrete-time signal is then converted from parallel to serial before passing through the digital-to-analog conversion to produce the following continuous-time domain transmit signal $s_{\text{CPP}}(t)$
\begin{IEEEeqnarray}{rCl}
	s_{\text{CPP}}(t) & = & 
	\begin{cases}
		s(t), & \hfill 0 \leq t \leq T, \\
		s(t + T) e^{-j 2\pi c_1(T^2 + 2 T t)/T_s^2}, & - L_{\text{cp}} T_s \leq t \leq 0,
	\end{cases} \nonumber\IEEEeqnarraynumspace
\end{IEEEeqnarray}
where  $T_s = 1/B$ is the sampling time, and $s(t)$ is defined as~\cite{bemani2024integrated}
\begin{IEEEeqnarray}{RCL}
	s(t) &{}={}& \frac{1}{\sqrt{T}} \sum_{m=0}^{N-1} x_m \, e^{j2\pi\left(c_{2}\,m^{2} \;+\;\frac{c_1}{T_s^2} t^2 + \frac{m}{T} t -\frac{q}{T_s} t\right)}, \nonumber \\ && \hfill 0 \le t \in [t_{m,q}, t_{m,q+1}]  < T, \: q = 0, ..., 2Nc_1,
\end{IEEEeqnarray}	
and 
\begin{IEEEeqnarray}{rCl}
	t_{m,q} & = & 
	\begin{cases}
		0, & q = 0, \\[10pt]
		\frac{(N - m)}{2 N c_1} T_s + \frac{q - 1}{2 c_1} T_s, & 1 \leq q \leq 2Nc_1.
	\end{cases}
\end{IEEEeqnarray}

The transmit signal then passes through the following doubly selective channel
\begin{IEEEeqnarray}{RCL}
	h(t, \tilde\tau) = \sum_{p = 1}^{P} h_p \delta(\tilde\tau - \tilde\tau_p) e^{-j 2\pi \tilde\nu_p t},
\end{IEEEeqnarray}
where $P$ is the number of resolvable paths, $\tilde\tau_p \geq 0$ and $\tilde\nu_p$ are the true delay and true Doppler shift of the $p$th path, measured in seconds and hertz, respectively, and $h_p$ is the complex channel gain of the $p$th path. The received signal in the continuous-time domain is given as
\begin{IEEEeqnarray}{RCL}
	r(t) &{}={}&  \sum_{p = 1}^{P} h_p e^{-j 2\pi \tilde\nu_p t} s_{\text{CPP}}(t - \tilde\tau_p) + w(t),
\end{IEEEeqnarray}	
where $w(t)$ is the complex additive white Gaussian noise with zero mean and one-sided power spectral density (PSD) of $N_0$. The received signal $r(t)$ is then sampled every $T_s$, converted from serial to parallel, and the CPP is removed before being applied to the  discrete affine Fourier transform (DAFT), whose output samples can be processed by any of the detectors in \cite{bemani2023affine} to reconstruct the transmitted $M$-ary QAM symbols.

\section{Analysis of the Ambiguity Function}\label{sec:AF}
One may observe the similarities between the AFDM waveform and other chirp-based waveforms, e.g., linear and nonlinear frequency-modulated chirps \cite{richards2005fundamentals}, used for various radar applications. However, the AFDM waveform is different, as it is the sum of multiple chirps, where each of them carries random $M$-ary QAM data symbols to also facilitate communication; while chirp-based signals used in radar typically carry known pre-designed sequences. 
That being said, for the AFDM waveform modulated with  $M$-ary QAM data symbols, we study its ambiguity function  which is the output of the matched filter and represents a two-dimensional correlation function between the transmit signal and a time-delayed and frequency-shifted version of itself \cite{richards2005fundamentals}. The complex-valued ambiguity function is defined as
\begin{IEEEeqnarray}{RCL}
	\tilde A_\text{u}(\tilde\tau, \tilde\nu)=\int_{0}^{T} s(t)\,s^*(t-\tilde\tau)\,e^{j2\pi \tilde\nu t} \text{d}t.
\end{IEEEeqnarray}
We normalize the true delay $\tilde\tau$ by the sampling time $T_s$ and the true Doppler $\tilde\nu$ by the chirp subcarrier spacing $\Delta_f$ to obtain the normalized delay and normalized Doppler frequency shift, respectively, as $\tau {}={} {\tilde\tau}/{T_s}$ and $\nu {}={} {\tilde\nu}/{\Delta_f}$.
Then, the complex-valued sampled ambiguity function of the normalized delay and normalized Doppler frequency shift, $A_\text{u}(\tau, \nu)$, is written as
\begin{IEEEeqnarray}{rcl}\label{eq:af}
	A_\text{u}(\tau, \nu)  &{}={}& 	\sum_{m,m'=0}^{N-1} x_m\,x^*_{m'}\,\sinc(\beta_{m,m'} (\tau, \nu)) e^{j\Psi_{m,m'}(\tau,\nu)}, \nonumber \\ && 
\end{IEEEeqnarray}	
where $\sinc(x) = \frac{\sin(\pi x)}{\pi x}$, 
\begin{IEEEeqnarray}{rcl}
	\beta_{m,m'}(\tau, \nu) &{}={}&  2c_1 N \tau + \nu + (m-m'), \\
	\Psi_{m,m'}(\tau,\nu) &{}={}&  2\pi\,\Phi_{m,m'}(\tau)+\pi\,{\beta}_{m,m'}(\tau,\nu),\\
	\Phi_{m,m'}(\tau) &{}={}& c_{2}(m^2-{m'}^2)-c_1\,\tau^2 + \left(\frac{m'}{N} -q\right) \tau. \IEEEeqnarraynumspace
\end{IEEEeqnarray}	
Since the maximum value of the ambiguity function is $A_\text{u}(0, 0)$ at which $\beta_{m,m'}(0, 0) = m - m'$ and $\Phi_{m,m'}(0) = c_{2}(m^2-{m'}^2)$, one can calculate $\E(A_\text{u}(0, 0)) {}={}  \sum_{m = 0}^{N-1} \sinc(0) \; e^{j 2 \pi 0} = N.$
Hence, we define the normalized complex-valued sampled ambiguity function as
\begin{IEEEeqnarray}{rcl}\label{eq:af_n}
	A(\tau, \nu) &{}={}& \frac{1}{N} A_\text{u}(\tau, \nu),
\end{IEEEeqnarray}
and in the rest of the paper we will refer to $A(\tau, \nu)$ as the ambiguity function for brevity. One can observe the coupling between the delay/range and the Doppler/velocity of the AFDM waveform in the $\sinc(\beta_{m,m'}(\tau, \nu))$ term, which suggests that the AFDM waveform is a Doppler tolerant waveform. 

To gain insights into the design of the AFDM waveform for sensing applications in the presence of random $M$-ary QAM data symbols, we study the effect of the AFDM parameters $c_1$ and $c_2$ on the zero-delay response, $A(0, \nu)$, and zero-Doppler response, $A(\tau, 0)$.

One can see from \eqref{eq:af_n} that $A(\tau, 0) = 0$ when $\beta_{m,m'}(\tau, 0)$ in an integer that does not equal zero, i.e., $\beta_{m,m'}(\tau, 0) \in \mathbb{Z} \backslash 0$. This is equivalent to
\begin{IEEEeqnarray}{rcl}\label{eq:sensing_condition_temp}
	2 c_1 N \tau \in \mathbb{Z} &{}\quad \text{and} \quad{}& 2 c_1 N \tau \neq (m - m'), \: \forall m, m'.
\end{IEEEeqnarray}
Please note that the second condition that $2 c_1 N \tau \neq (m - m'), \: \forall m, m'$ does not appear in the derivation in \cite{zhu2024afdm} as the authors considered a known symbols of value 1 to be carried by the AFDM waveform,  and they did not consider the general case of having the AFDM waveform carrying $M$-ary QAM data symbols.
Since $(m - m') \in \{-(N-1), ..., (N-1)\}$, the condition in \eqref{eq:sensing_condition_temp} is re-written as
\begin{eqnarray}\label{eq:sensing_condition}
	2 c_1 N \tau \in \mathbb{Z}_{\geq N},
\end{eqnarray}
where $\mathbb{Z}_{\geq N}$ denotes the set of integer numbers greater than or equal to $N$, i.e.,  $\mathbb{Z}_{\geq N}= \{z\in\mathbb{Z}\mid z\geq N\}$. 
We refer to \eqref{eq:sensing_condition} as the sensing condition on the chirp rate $c_1$ to  obtain the maximum delay/range resolution in the presence of random $M$-ary QAM symbols. Please note that the sensing condition on $c_1$ in \eqref{eq:sensing_condition} imposes an additional constraint on the optimal diversity order condition of the chirp rate $c_1$ that only requires $2 c_1 N$ to be an integer and not necessarily $\geq N$.

{For the zero-delay response, i.e., $A(0, \nu)$, one can see that neither $c_1$ nor $c_2$ appears in  $\beta_{m,m'}( 0, \nu)$, and apparently the AFDM parameters cannot be designed to reduce the sidelobes in the Doppler domain in the presence of random $M$-ary QAM.}

\section{Approximate Distribution for the Magnitude of the Ambiguity Function}\label{sec:proposed}
In evaluating the sensing performance of a given waveform, we study the characteristics of the magnitude of the ambiguity function, i.e., $|A(\tau, \nu)|$. 
However, the ambiguity function in \eqref{eq:af} depends on the random $M$-ary QAM data symbols, and hence, analyzing the sensing performance of the AFDM will also depend on the random $M$-ary QAM data symbols. Hence, to facilitate studying and quantifying the sensing performance of the AFDM waveform, in this section we find an approximate distribution of the magnitude of the ambiguity function, i.e., $|A(\tau, \nu)|$.

One can see from \eqref{eq:af} that the ambiguity function is a summation of $N^2$ terms that include  the random variables  $x_mx^*_{m'}x_{m'}x^*_m,$ where $m, m' = 0, ..., N-1$. Please note that not all the $N^2$ terms are independent; for example, $x_m x^*_{m'}x_{m'}x^*_m$ and $x_mx^*_{k'}x_{m'}x^*_k$ share $x_mx_{m'}$. However, the correlation between different terms decays due to $\sinc({\beta}_{m,m'}(\tau,\nu))$ and the dominant terms may be weakly correlated or uncorrelated especially when the index difference $m - m'$ increases, and  the central limit theorem often holds in such scenarios. Accordingly, we can approximate the ambiguity function as
\begin{IEEEeqnarray}{RCL}
	A(\tau, \nu) \sim \mathcal{CN}(\mu_\text{A}(\tau, \nu), \sigma^2_\text{A}(\tau, \nu)),
\end{IEEEeqnarray}
where $\mu_\text{A}(\tau, \nu)$ and $\sigma^2_\text{A}(\tau, \nu)$ are the mean and the variance of the approximate complex Gaussian  distribution. In the following, we derive $\mu_\text{A}(\tau, \nu)$ and $\sigma^2_\text{A}(\tau, \nu)$.

The ambiguity function expression in \eqref{eq:af} can be written in a compact form as
\begin{IEEEeqnarray}{RCL}\label{eq:af_compact}
	A(\tau,\nu) &{}={}& \frac{1}{N} \, \mathbf{x}^H\,\mathbf{W}(\tau,\nu)\,\mathbf{x}, \\
	&{}={}&  \frac{1}{N} \, \tr\left( \mathbf{W}(\tau,\nu)\,\mathbf{x}\mathbf{x}^H\right),
\end{IEEEeqnarray}
where $\mathbf{x} = \left[x[0], \dots, x[N-1] \right]^{\text{T}} \in \mathbb{C}^{N \times 1}$ is the vector of all $M$-ary QAM symbols and  $\mathbf{W}(\tau,\nu) \in \mathbb{C}^{N \times N}$ whose $(m,m')$ element is defined as
\begin{IEEEeqnarray}{rcl}
	\mathbf{W}_{m,m'}(\tau,\nu)  &{}={}& \sinc(\beta_{m,m'} (\tau, \nu)) e^{j\Psi_{m,m'}(\tau,\nu)}.
\end{IEEEeqnarray}	
By using the linearity of the trace and  expectation operators, one can show that the mean of $A(\tau, \nu)$ can be evaluated as 
\begin{IEEEeqnarray}{RCL}
	\mu_\text{A}(\tau, \nu) &{}={}& \E\left(A(\tau, \nu)\right) = \frac{1}{N} \tr\left(\mathbf{W}(\tau,\nu)\right),
\end{IEEEeqnarray}
where $\mathbb{E}(\mathbf{x}\mathbf{x}^H) = \mathbf{I}$, i.e., the $M$-ary QAM symbols are normalized to have unit power. 

To find $\sigma^2_\text{A}(\tau, \nu) = \V(A(\tau, \nu))$, one has to find the second moment, i.e., $\E(A(\tau, \nu) A^*(\tau, \nu)) = \E(|A(\tau, \nu)|^2)$ first, and this is what we do next. The expectation of the square of the magnitude of the ambiguity function can be written, with the help of \eqref{eq:af_compact},  as
\begin{IEEEeqnarray}{RCL} \label{eq:sq_abs}
	\E(|A(\tau,\nu)|^2)
	&{}={}& \frac{1}{N^2} \sum_{i,j,k,\ell} \E\left(x_i^*\,W_{ij}(\tau,\nu)\,x_j\; x_k\,W^*_{k \ell}(\tau,\nu)\,x_\ell^*\right).  \nonumber \\ 
\end{IEEEeqnarray}
Taking the expectation of 	$|A(\tau,\nu)|^2$ requires evaluating $\E(x_i^* x_j x_k x_\ell^*)$ which can be evaluated as \cite{mendel1991tutorial}
\begin{IEEEeqnarray}{rcl}\label{eq:fourth_moment}
	\E(x_i^* x_j x_k x_\ell^*) &{}={}& \E(x_i^* x_j) \E(x_k x_\ell^*) +  \E(x_i^*  x_k) \E(x_j x_\ell^*) \nonumber \\
	&{}{}& + \E(x_i^* x_\ell^*) \E(x_j x_k) + \kappa_4(x_i^*, x_j, x_k, x_\ell^*), \IEEEeqnarraynumspace
\end{IEEEeqnarray}
where $\kappa_4(x_i^*, x_j, x_k, x_\ell^*)$ is the fourth-order joint cumulant that accounts for the non Gaussian characteristics of the $M$-ary QAM, i.e., $\kappa_4(x_i^* x_j x_k x_\ell^*) \neq 0$ for $M$-ary QAM data symbols.

Recall that we consider proper $M$-ary QAM symbols, i.e., $\mathbb{E}\{x_i\}=0$ and $\mathbb{E}\{x_i^2\}=0$, with unit power, i.e, $\E\left(|x_i|^2\right) = 1$; hence, to evaluate \eqref{eq:fourth_moment} we consider the following five cases.

-- Case 1 ($i = j = k = \ell$): In this case, we evaluate \eqref{eq:fourth_moment} as
\begin{IEEEeqnarray}{rcl}\label{eq:case1}
	\E(|x_i|^4) &{}={}& 2 + \kappa_4(x_i^*, x_i, x_i, x_i^*). 
\end{IEEEeqnarray}
Please note that for unit power proper $M$-ary QAM, $\E(|x_i|^4)$ will be the kurtosis of the constellation and its value for square $M$-ary QAM can be approximated as $(7M - 13)/(5M - 5)$.

-- Case 2 ($i \neq j \neq k \neq \ell$): We calculate \eqref{eq:fourth_moment} as
\begin{IEEEeqnarray}{rcl}\label{eq:case2}
	\E(x_i^* x_j x_k x_\ell^*) &{}={}& 0,
\end{IEEEeqnarray}
where $\kappa_4(x_i^*, x_j, x_k, x_\ell^*) = 0$ for independent and proper $M$-ary QAM symbol.

-- Case 3 ($i = j, k = \ell, j \neq k$): We evaluate \eqref{eq:fourth_moment} as
\begin{IEEEeqnarray}{rcl}\label{eq:case3}
	\E(x_i^* x_i x_k x_k^*) &{}={}& 1.
\end{IEEEeqnarray}

-- Case 4 ($i = k, j = \ell, j \neq k$): Similar to Case 3, we calculate \eqref{eq:fourth_moment} as
\begin{IEEEeqnarray}{rcl}\label{eq:case4}
	\E(x_i^* x_j x_i x_j^*) &{}={}& 1.
\end{IEEEeqnarray}

-- Case 5 ($i = \ell, j = k, j \neq \ell$): Similar to Case 2, we evaluate \eqref{eq:fourth_moment} as
\begin{IEEEeqnarray}{rcl}\label{eq:case5}
	\E(x_i^* x_j x_j x_i^*) &{}={}& 0.
\end{IEEEeqnarray}
Accordingly, one can re-express \eqref{eq:fourth_moment} as
\begin{IEEEeqnarray}{rcl}\label{eq:fourth_moment_final}
	\mathbb{E} (x_i^*\,x_j\,x_k\,x_\ell^*) &{}={}& \delta_{ij}\,\delta_{k\ell} + \delta_{ik}\,\delta_{j\ell} \nonumber \\ &{} & + (\E(|x_i|^4) - 2)\,\delta_{ij}\,\delta_{jk}\,\delta_{k\ell}. \IEEEeqnarraynumspace
\end{IEEEeqnarray}
Hence, we re-write the second moment in \eqref{eq:sq_abs} as
\begin{IEEEeqnarray}{rcl}
	\E(|A(\tau,\nu)|^2) &{}={}& \frac{1}{N^2}\sum_{i,k} \mathbf{W}_{ii}(\tau,\nu)\,\mathbf{W}^*_{kk}(\tau,\nu) \nonumber \\ && + \frac{1}{N^2}\sum_{i,j} \mathbf{W}_{ij}(\tau,\nu)\,\mathbf{W}^*_{ij}(\tau,\nu) \nonumber \\ && + \frac{(\E(|x_i|^4) - 2) }{N^2}\sum_i \mathbf{W}_{ii}(\tau,\nu)\,\mathbf{W}^*_{ii}(\tau,\nu), \IEEEeqnarraynumspace
\end{IEEEeqnarray}	
which can be further re-express in a compact form as
\begin{IEEEeqnarray}{rcl}
	\E(|A(\tau,\nu)|^2) &{}={}& \mu^2_\text{A} + \frac{1}{N^2}\tr\Bigl(\mathbf{W}(\tau,\nu)\,\mathbf{W}(\tau,\nu)^\text{H}\Bigr) \nonumber \\ &&
	+ \frac{(\E(|x_i|^4) - 2)}{N^2}   \|\mathbf{w}(\tau,\nu)\|_2^2, 
\end{IEEEeqnarray}
where $\mathbf{w}(\tau,\nu) = \operatorname{diag}\big(\mathbf{W}(\tau,\nu)\big)$. 
Hence, the variance of $A(\tau, \nu)$ is calculated as
\begin{IEEEeqnarray}{rcl}\label{eq:var}
	\sigma^2_\text{A}(\tau,\nu) &{}={}& \frac{1}{N^2}\tr\Bigl(\mathbf{W}(\tau,\nu)\,\mathbf{W}(\tau,\nu)^\text{H}\Bigr) \nonumber \\ && +  \frac{(\E(|x_i|^4) - 2)}{N^2}  \|\mathbf{w}(\tau,\nu)\|_2^2.  \IEEEeqnarraynumspace
\end{IEEEeqnarray}

Given $\mu_\text{A}(\tau,\nu)$ and $\sigma^2_\text{A}(\tau,\nu)$ of the approximate complex Gaussian distribution of $A(\tau, \nu)$, it follows directly that $|A(\tau, \nu)|$ has a Rice distribution as
\begin{IEEEeqnarray}{RCL}
	|A(\tau, \nu)| \sim \mathcal{R}(s_\text{R}(\tau,\nu), \sigma^2_\text{R}(\tau,\nu)),
\end{IEEEeqnarray}
where $s_\text{R}(\tau,\nu) = |\mu_\text{A}(\tau,\nu)|$ and $\sigma^2_\text{R}(\tau,\nu) = \frac{1}{2} \sigma^2_\text{A}(\tau,\nu)$ are the parameters of the Rice distribution. 

\section{Evaluation of the PSLR and ISLR}\label{sec:PSLR_ISLR}
We use the PSLR to quantify how well the AFDM waveform can suppress the sidelobes compared to its mainlobe. It is formally defined as \cite{weinberg2018topics}
\begin{IEEEeqnarray}{rcl} \label{eq:PSLR}
	\text{PSLR} = 20\log_{10}\left(\frac{|A_\text{u}(\breve{\tau}, \breve{\nu})|}{|A_\text{u}(0, 0)|}\right) = 20\log_{10}\left(|A(\breve{\tau}, \breve{\nu})|\right), \IEEEeqnarraynumspace
\end{IEEEeqnarray} 
where $(\breve{\tau}, \breve{\nu})$ denotes the location of the largest sidelobe. A lower PSLR value indicates that the highest sidelobe is much lower than the mainlobe, which reduces the chance of false alarms due to strong sidelobe responses.

Additionally, we use the ISLR to quantify the energy in the sidelobes of the AFDM waveform when compared to the energy in its mainlobe. It is formally defined as \cite{weinberg2018topics}
\begin{IEEEeqnarray}{rcl}\label{eq:ISLR}
	\text{ISLR} &{}={}& 10 \log_{10} \left( \frac{\displaystyle \iint_{(\tau, \nu) \in \mathcal{S}} \left| A(\tau, \nu) \right|^2 \text{d}\tau\, \text{d} \nu}{\displaystyle \iint_{(\tau, \nu) \in \mathcal{M}} \left| A(\tau, \nu) \right|^2 \text{d}\tau\, \text{d} \nu} \right), \IEEEeqnarraynumspace
\end{IEEEeqnarray}
where $\mathcal{S}$ and $\mathcal{M}$ are the set of $(\tau, \nu)$ points corresponding to the sidelobes and mainlobe, respectively. Lower ISLR values indicate that the overall sidelobes energy relative to the mainlobe energy is small, which contributes to improved target detection.

As can be seen from \eqref{eq:PSLR} and \eqref{eq:ISLR}, both the PSLR and the ISLR depend on the unknown $M$-ary QAM symbols. To facilitate the calculation of the PSLR and ISLR, we evaluate both \eqref{eq:PSLR} and \eqref{eq:ISLR} after averaging the ambiguity function $A(\tau, \nu)$ in \eqref{eq:af_n} over the unknown $M$-ary QAM symbols. That being said, one can show that
\begin{IEEEeqnarray}{rcl}
	\E(A(\tau, \nu)) &{}={}& \frac{1}{N} 	\sum_{m=0}^{N-1} \sinc(\beta_{m,m} (\tau, \nu)) e^{j\Psi_{m,m}(\tau,\nu)},
\end{IEEEeqnarray}
and since $\beta_{m,m} = 2 c_1 N \tau + \nu$ does not depend on $m$, one can write the magnitude of the expected value of the ambiguity function as
\begin{IEEEeqnarray}{rcl}\label{eq:abs_E_AF}
	|\E(A(\tau, \nu))| &{}={}& \frac{1}{N}|\sinc(2 c_1 N \tau + \nu)| \frac{|\sin(\pi \tau)|}{|\sin(\pi \tau /N)|}.
\end{IEEEeqnarray}
From \eqref{eq:abs_E_AF}, one can identify the location of the points $(\tau,\nu)$ of the sets $\mathcal{M}$ and $\mathcal{S}$. For example, to identify the set $\mathcal{M}$ of the mainlobe, one need to find the vertices of the parallelogram containing the mainlobe as follows. The term $\sinc(2 c_1 N \tau + \nu)$ in \eqref{eq:abs_E_AF} has its first zeros at $2 c_1 N \tau + \nu = \pm 1$. Similarly, the term $ \frac{|\sin(\pi \tau)|}{|\sin(\pi \tau /N)|}$ has its first zeros at $\tau = \pm 1$. Hence, the four vertices of the parallelogram containing the mainlobe are: $(1, 1-2c_1 N)$, $(1, -1-2c_1 N)$, $(-1, 2 c_1 N + 1)$, $(-1, 2 c_1 N - 1)$.

\section{Simulation Results}\label{sec:simulation}
In this section, we evaluate the ambiguity function  of  AFDM   and also evaluate the accuracy of its approximate  distribution derived in Section \ref{sec:proposed}. We additionally compute the PSLR and ISLR obtained from the evaluation of the ambiguity function and its approximate distribution. The adopted simulation parameters are as follows. The AFDM chirp subcarriers is set to 64, 128, and 256. The carrier frequency is 24 GHz, the bandwidth is 93.1 MHz, and the constellation size is set to 4, 16, 64, and 256. The results are averaged over 1000 iterations. Unless otherwise mentioned, we set the chirp rate $c_1 = 1$ according to \eqref{eq:sensing_condition} and $c_2 = 0$. The results are presented for a wide range of normalized Doppler values to accommodate simulation parameters beyond those used in the paper. 

\begin{figure}[t]
	\centering
	{\includegraphics[width=0.33\textwidth]{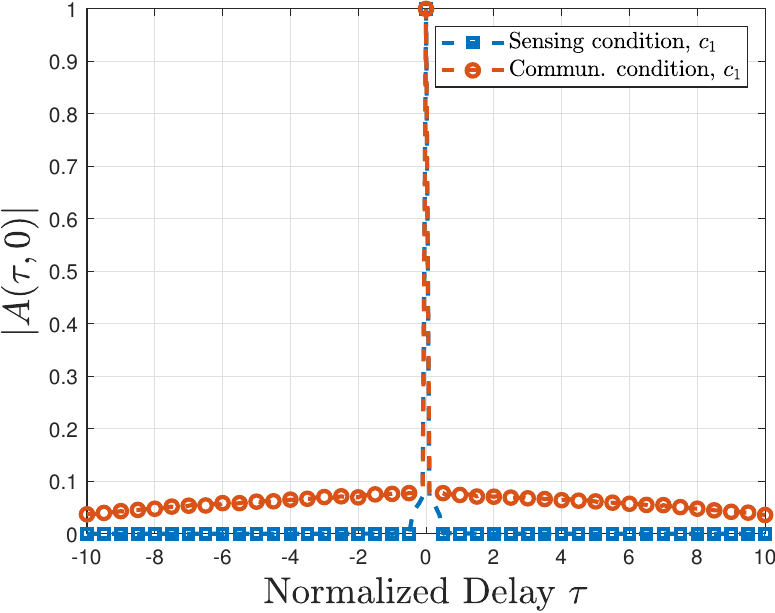}}
	\caption{Zero-Doppler response for $N$ = 128 and $M$ = 4.}
	\label{fig:effect_c1}
\end{figure}

To illustrate the effect of the chirp rate $c_1$ on the sensing performance of the AFDM waveform, Fig. \ref{fig:effect_c1} shows the zero-Doppler response for two cases: 1) selecting the chirp rate $c_1$ to achieve the optimal diversity order as per \cite{bemani2023affine} or good sensing performance when the modulation symbols are fixed and have a value of 1 as per \cite{zhu2024afdm}, and 2) selecting the chirp rate $c_1$ to achieve the optimal sensing performance in the presence of unknown $M$-ary QAM data symbols, as derived in this paper in \eqref{eq:sensing_condition}. It is evident from Fig. \ref{fig:effect_c1} that choosing the chirp rate $c_1$ according to \eqref{eq:sensing_condition} significantly reduces the sidelobes in the delay/range domain, thereby improving the delay/range resolution. Selecting the chirp rate $c_1$ according to \cite{bemani2023affine, zhu2024afdm} will result in a much higher level of sidelobes.

\begin{figure}[t]
	\centering
	\subfloat[Empirical average]{\includegraphics[width=0.24\textwidth]{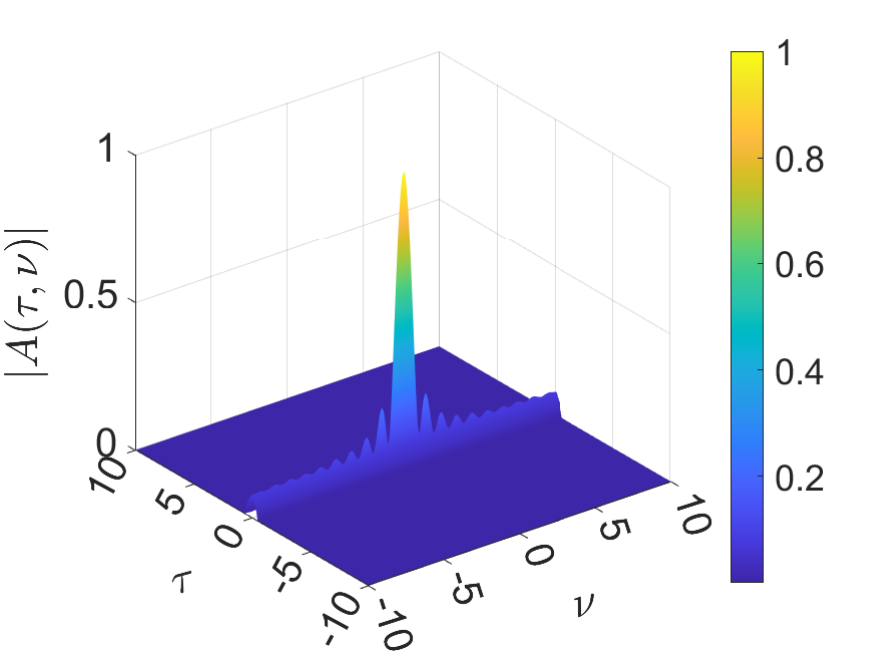}\label{fig:empirical_abs_A}}
	\hfill
	\subfloat[Approximate Rice distribution]{\includegraphics[width=0.24\textwidth]{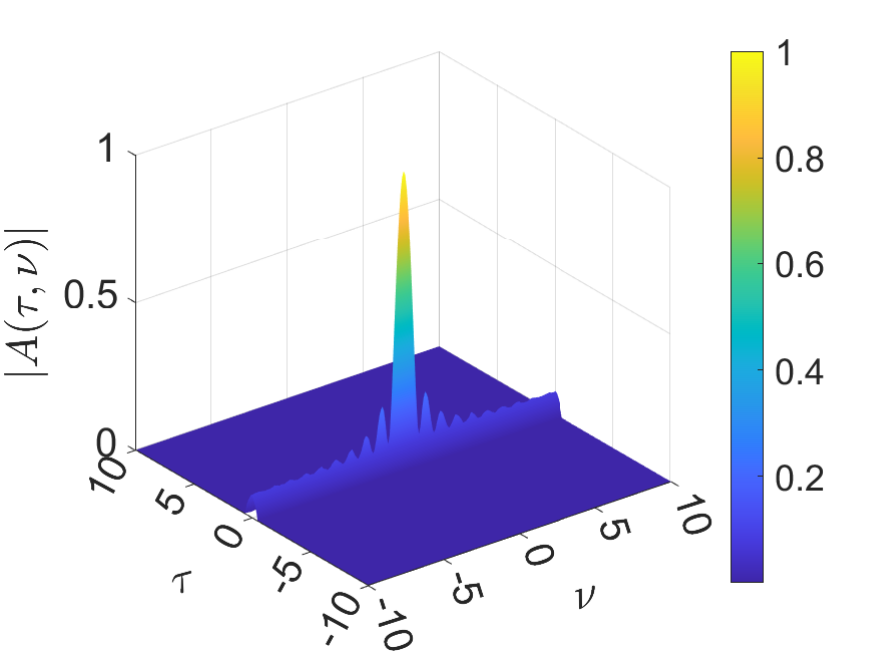}\label{fig:approx_abs_A}}
	\caption{Comparison of empirical average and Rice distribution of $|A(\tau,\nu)|$ for $N$ = 128 and $M$ = 4.}
	\label{fig:3D}
\end{figure}

Fig. \ref{fig:empirical_abs_A} shows the empirical average of the magnitude of the ambiguity function, i.e., $|A(\tau,\nu)|$, over 1000 trials for $N = 128$ and $M = 4$, as a function of the normalized delay $\tau$ and normalized Doppler $\nu$. The approximate Rice distribution of $|A(\tau,\nu)|$ is presented in Fig. \ref{fig:approx_abs_A}, and it closely matches the empirical average of $|A(\tau,\nu)|$. This matching will be further verified through the PSLR and ISLR values.

\begin{figure}[t]
	\centering
	\subfloat[Empirical average]{\includegraphics[width=0.24\textwidth]{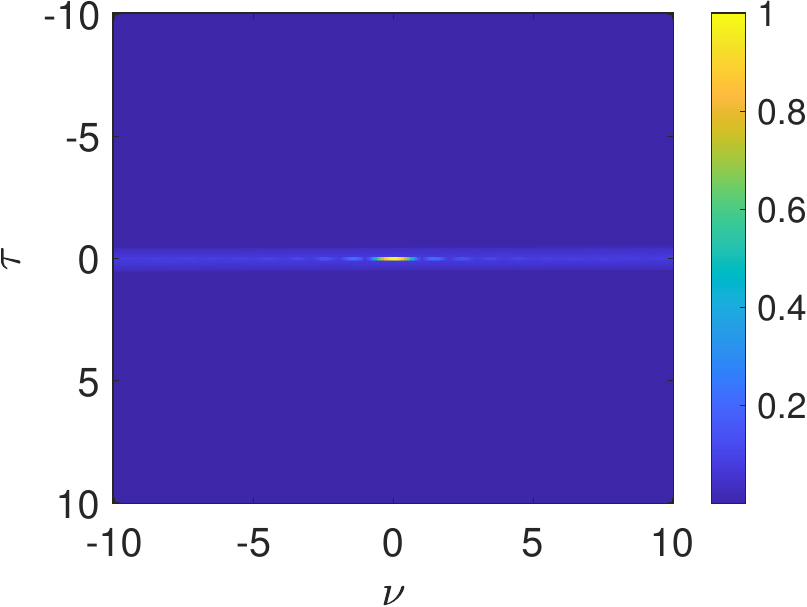}\label{fig:empirical_abs_A_2D}}
	\hfill
	\subfloat[Approximate Rice distribution]{\includegraphics[width=0.24\textwidth]{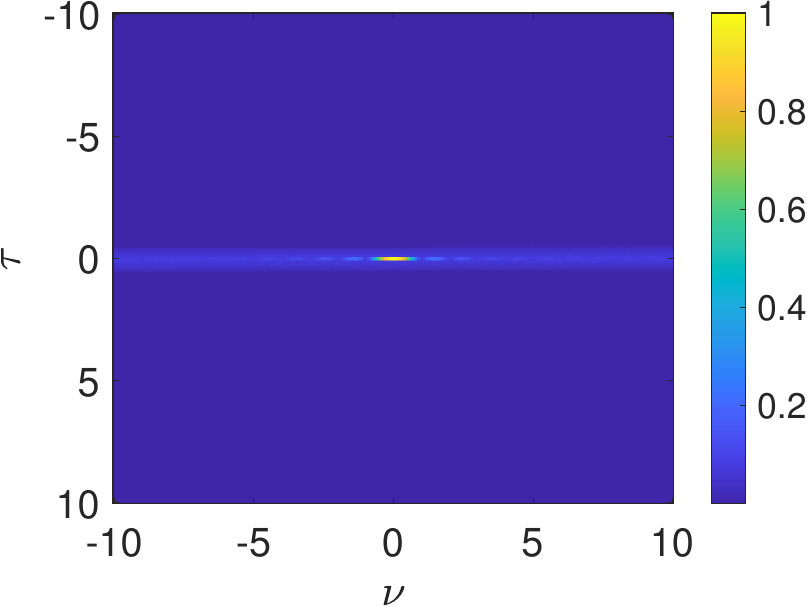}\label{fig:approx_abs_A_2D}}
	\caption{Comparison of empirical average and Rice distribution of the delay-Doppler map of $|A(\tau,\nu)|$ for $N$ = 128 and $M$ = 4.}
	\label{fig:2D}
\end{figure}

The delay-Doppler maps of the empirical average of $|A(\tau,\nu)|$ and its approximate Rice distribution are shown in Figs. \ref{fig:empirical_abs_A_2D} and \ref{fig:approx_abs_A_2D}, respectively. As observed, the AFDM waveform achieves excellent delay resolution when the chirp rate $c_1$ is chosen according to \eqref{eq:sensing_condition}. Additionally, one can confirm that $|A(\tau,\nu)|$ can be well approximated by the Rice distribution.

\begin{figure}[t]
	\centering
	\subfloat[3D]{\includegraphics[width=0.24\textwidth]{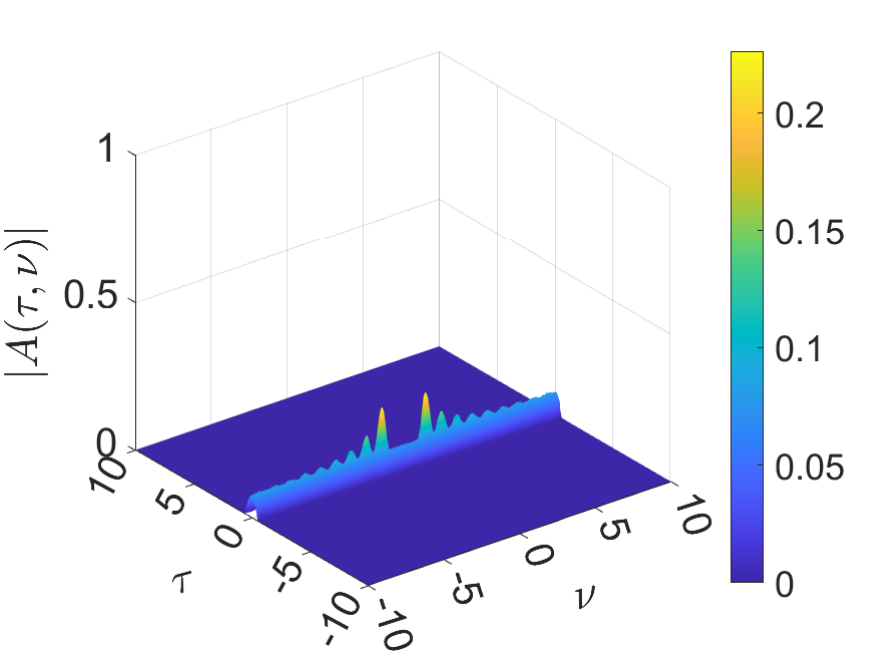}}
	\hfill
	\subfloat[Delay-Doppler map]{\includegraphics[width=0.24\textwidth]{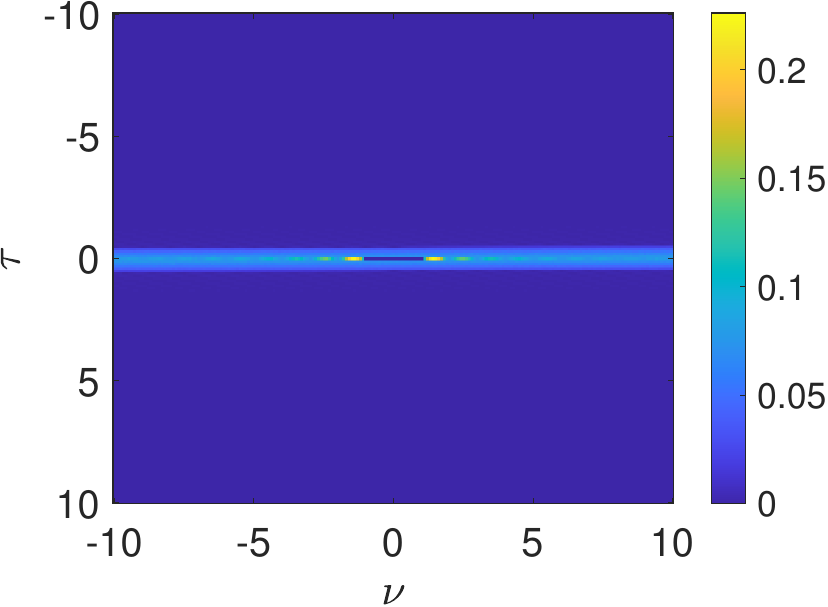}}
	\caption{$|A(\tau,\nu)|$ without the mainlobe for $c_1 = 1$, $N = 128$, and $M = 4$.}
	\label{fig:mainlobe_1}
\end{figure}

\begin{figure}[t]
	\centering
	\subfloat[3D]{\includegraphics[width=0.24\textwidth]{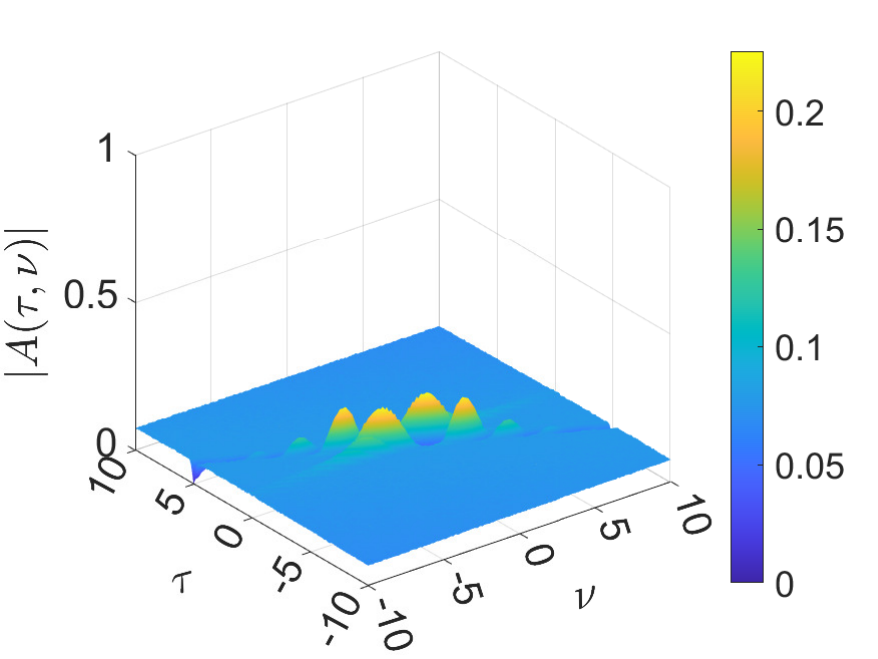}}
	\hfill
	\subfloat[Delay-Doppler map]{\includegraphics[width=0.24\textwidth]{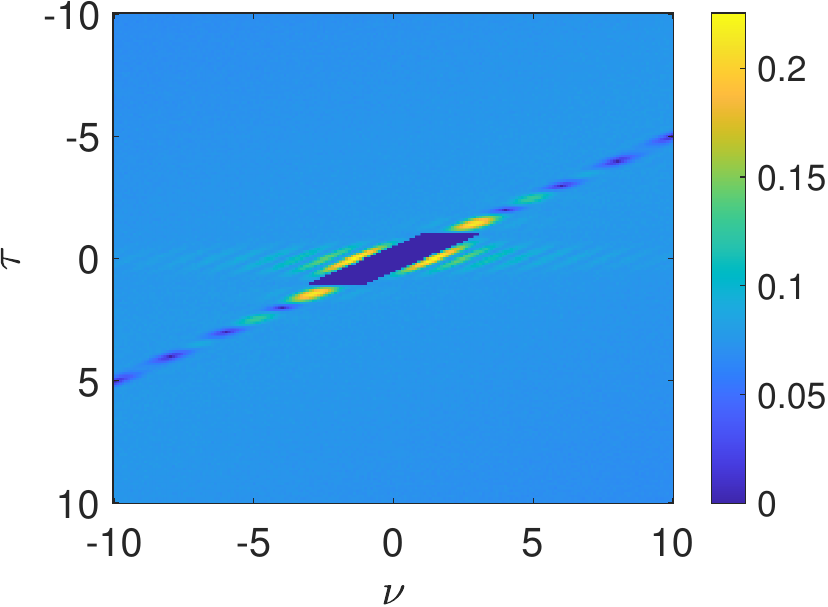}}
	\caption{$|A(\tau,\nu)|$ without the mainlobe for $c_1 = 1/N$, $N = 128$, and $M = 4$.}
	\label{fig:mainlobe_2}
\end{figure}

To calculate of the PSLR and ISLR, we compute the boundaries of the mainlobe and sidelobes. Fig. \ref{fig:mainlobe_1} and Fig. \ref{fig:mainlobe_2} shows $|A(\tau,\nu)|$ after identifying and removing the parallelogram of the mainlobe for $c_1 = 1$ and $c_1 = 1/N$, respectively, for $N = 128$ and $M = 4$. As can be seen, the calculated vertices of the mainlobe as given in Section \ref{sec:PSLR_ISLR} can accurately identify the mainlobe. Please note that selecting the chirp rate $c_1$ according to \eqref{eq:sensing_condition} results in a much lower sidelobe levels (see Fig. \ref{fig:mainlobe_1}) when compared to selecting $c_1$ as in  \cite{bemani2023affine, zhu2024afdm} (see Fig. \ref{fig:mainlobe_2}).

\begin{figure}[t]
	\centering
	\subfloat[PSLR (dB)]{\includegraphics[width=0.24\textwidth]{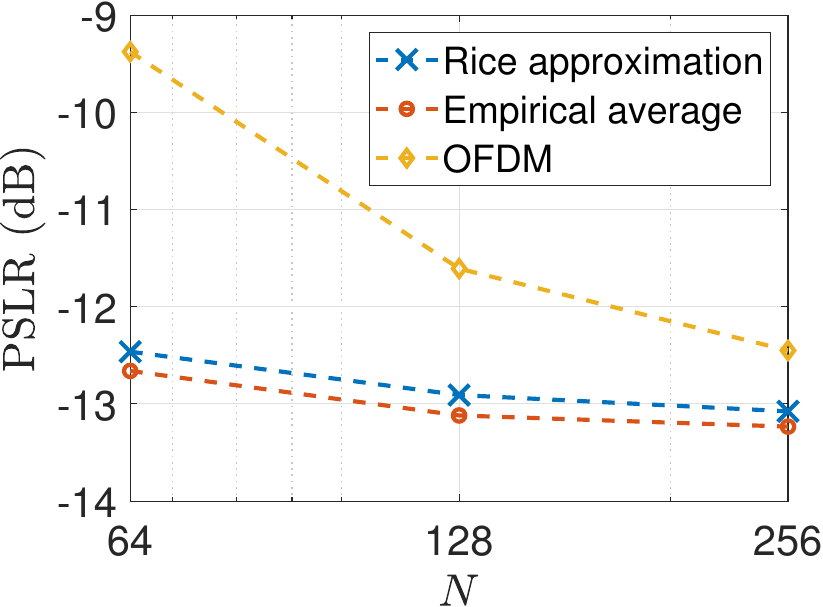}\label{fig:PSLR}}
	\hfill
	\subfloat[ISLR (dB)]{\includegraphics[width=0.24\textwidth]{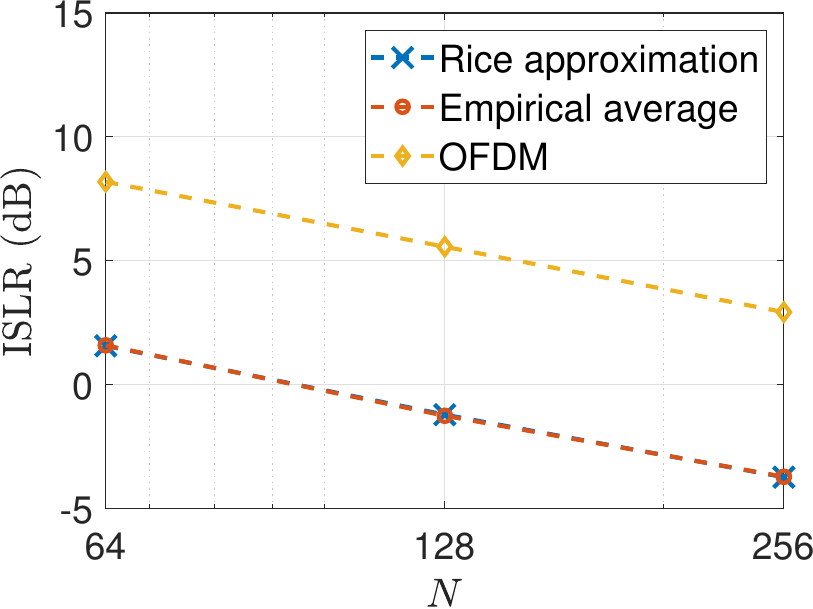}\label{fig:ISLR}}
	\caption{PSLR  and ISLR for the empirical average of $|A(\tau,\nu)|$ of AFDM, its approximate distribution, and  OFDM at $M$ = 4.}
	\label{fig:side_by_side_N}
\end{figure}

\begin{figure}[t]
	\centering
	\subfloat[PSLR (dB)]{\includegraphics[width=0.24\textwidth]{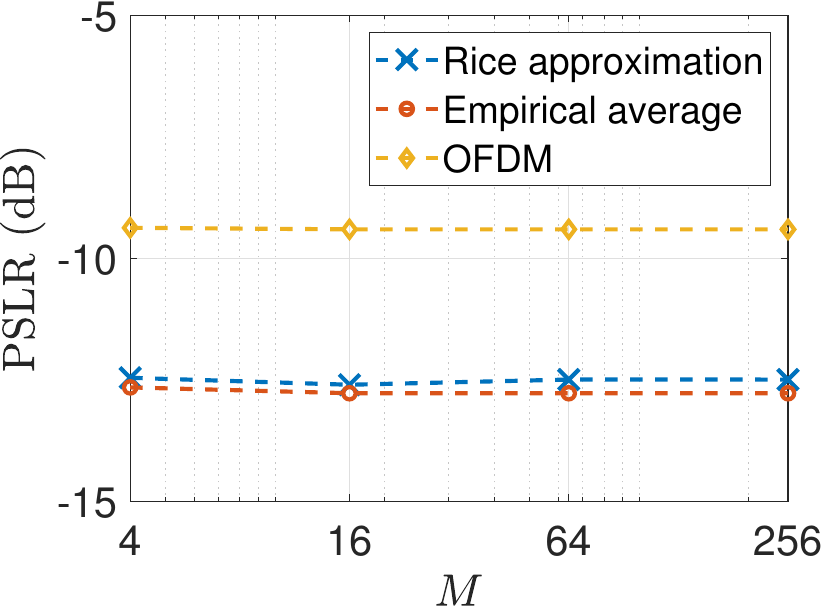}\label{fig:PSLR_M}}
	\hfill
	\subfloat[ISLR (dB)]{\includegraphics[width=0.24\textwidth]{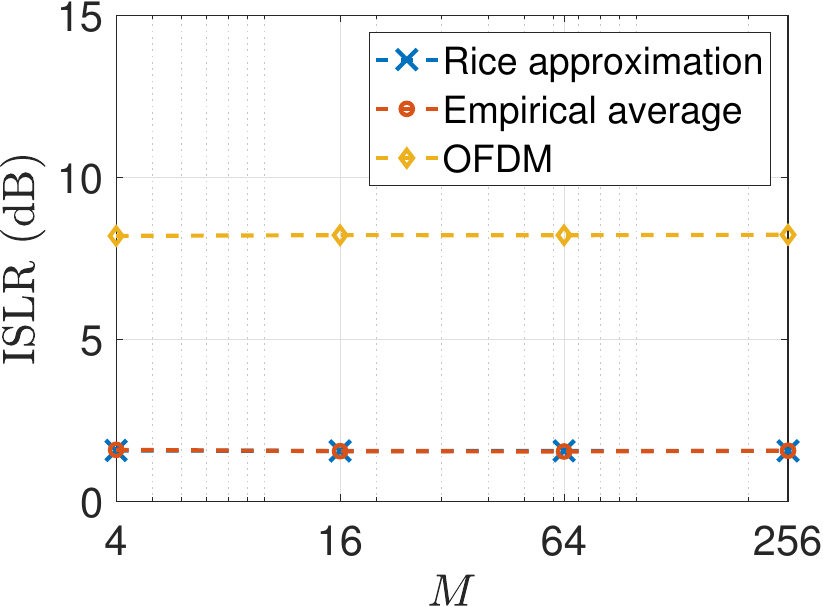}\label{fig:ISLR_M}}
	\caption{PSLR and ISLR for the empirical average of $|A(\tau,\nu)|$ of AFDM, its approximate distribution, and  OFDM at  $N = 64$.}
	\label{fig:side_by_side_M}
\end{figure}

The PSLR and the ISLR of the AFDM waveform are depicted in Fig. \ref{fig:PSLR} and Fig. \ref{fig:ISLR}, respectively, for different values of $N$ at $M = 4$. In particular, we plot the PSLR and ISLR values for the empirical average of $|A(\tau,\nu)|$ and its approximate Rice distribution. As can be seen, for a fixed total bandwidth, increasing the number of chirps improves both the PSLR and ISLR and improves the sensing performance. Additionally, Fig. \ref{fig:PSLR} and Fig. \ref{fig:ISLR} shows that the Rice distribution approximates the empirical $|A(\tau,\nu)|$ well. One can additionally see from Fig. \ref{fig:PSLR} and Fig. \ref{fig:ISLR} the superiority of the AFDM waveform when compared to the OFDM waveform as it achieves much lower PSLR and ISLR values for the same bandwidth. Please note that the OFDM waveform can be obtained from the AFDM waveform by setting $c_1 = c_2 = 0$.

The PSLR and ISLR for both the empirical average of $|A(\tau,\nu)|$ and its approximate Rice distribution of the AFDM waveform are shown in Fig. \ref{fig:PSLR_M} and Fig. \ref{fig:ISLR_M}, respectively, for various values of $M$ at $N = 64$. As observed, changing the modulation order $M$ has negligible effect on the PSLR and ISLR. This is expected as the modulation order $M$ slightly affects the variance of the ambiguity function through its kurtosis as can be seen in \eqref{eq:var}. Similar to Fig. \ref{fig:side_by_side_N}, once can see from Fig. \ref{fig:side_by_side_M} that the sensing performance of the AFDM waveform surpasses its counterpart of OFDM as it achieves lower PSLR and ISLR.

\section{Conclusions} \label{sec:conclu}
In this paper, we derived a closed-form expression for the ambiguity function  of AFDM waveforms. We additionally found the optimal value of the AFDM waveform chirp rate $c_1$ that maximizes the delay/range resolution in the presence of unknown $M$-ary QAM data symbols, and hence, improve the sensing performance. We additionally showed that the Rice distribution can approximate the magnitude of the ambiguity function well. Simulation results verified the accuracy of the approximate Rice distribution. Additionally, the results showed that, for a fixed total bandwidth, increasing the number of chirps $N$ can reduce both the PSLR and ISLR, while changing the modulation order $M$ has negligible effect on both the PSLR and ISLR. The results also showed the superiority of the AFDM waveform when compared to the OFDM waveform for sensing applications for the same bandwidth.

\bibliographystyle{IEEEtran}
\bibliography{IEEEabrv,References}

\begin{thebibliography}{10}
\providecommand{\url}[1]{#1}
\csname url@samestyle\endcsname
\providecommand{\newblock}{\relax}
\providecommand{\bibinfo}[2]{#2}
\providecommand{\BIBentrySTDinterwordspacing}{\spaceskip=0pt\relax}
\providecommand{\BIBentryALTinterwordstretchfactor}{4}
\providecommand{\BIBentryALTinterwordspacing}{\spaceskip=\fontdimen2\font plus
\BIBentryALTinterwordstretchfactor\fontdimen3\font minus
  \fontdimen4\font\relax}
\providecommand{\BIBforeignlanguage}[2]{{%
\expandafter\ifx\csname l@#1\endcsname\relax
\typeout{** WARNING: IEEEtran.bst: No hyphenation pattern has been}%
\typeout{** loaded for the language `#1'. Using the pattern for}%
\typeout{** the default language instead.}%
\else
\language=\csname l@#1\endcsname
\fi
#2}}
\providecommand{\BIBdecl}{\relax}
\BIBdecl

\bibitem{gonzalez2024integrated}
N.~G.-P. et~al., ``{The integrated sensing and communication revolution for 6G:
  Vision, techniques, and applications},'' \emph{Proc. {IEEE}}, vol. 112,
  no.~7, pp. 278--305, July 2024.

\bibitem{zhou2022integrated}
W.~Zhou, R.~Zhang, G.~Chen, and W.~Wu, ``Integrated sensing and communication
  waveform design: A survey,'' \emph{IEEE Open J. Commun. Soc.}, vol.~3, pp.
  1930--1949, 2022.

\bibitem{bemani2023affine}
A.~Bemani, N.~Ksairi, and M.~Kountouris, ``Affine frequency division
  multiplexing for next generation wireless communications,'' \emph{{IEEE}
  Trans. Wireless Commun.}, vol.~22, no.~11, pp. 8214--8229, Nov. 2023.

\bibitem{ni2022afdm}
Y.~Ni, Z.~Wang, P.~Yuan, and Q.~Huang, ``{An AFDM-based integrated sensing and
  communications},'' in \emph{Proc. International Symposium on Wireless
  Communication Systems}, 2022, pp. 1--6.

\bibitem{bemani2024integrated}
A.~Bemani, N.~Ksairi, and M.~Kountouris, ``Integrated sensing and
  communications with affine frequency division multiplexing,'' \emph{{IEEE}
  Wireless Commun. Lett.}, vol.~13, no.~5, pp. 1201--1221, May 2024.

\bibitem{zhu2024afdm}
J.~Zhu, Y.~Tang, F.~Liu, X.~Zhang, H.~Yin, and Y.~Zhou, ``{AFDM-based bistatic
  integrated sensing and communication in static scatterer environments},''
  \emph{{IEEE} Wireless Commun. Lett.}, vol.~13, no.~8, pp. 2245--2249, Aug.
  2024.

\bibitem{berggren2022joint}
F.~Berggren and B.~M. Popovi{\'c}, ``Joint radar and communications with
  multicarrier chirp-based waveform,'' \emph{IEEE Open J. Commun. Soc.},
  vol.~3, pp. 1702--1718, 2022.

\bibitem{richards2005fundamentals}
M.~A. Richards \emph{et~al.}, \emph{Fundamentals of Radar Signal
  Processing}.\hskip 1em plus 0.5em minus 0.4em\relax New York: Mcgraw-Hill,
  2005.

\bibitem{mendel1991tutorial}
J.~M. Mendel, ``Tutorial on higher-order statistics (spectra) in signal
  processing and system theory: Theoretical results and some applications,''
  \emph{Proc. {IEEE}}, vol.~79, no.~3, pp. 278--305, Mar. 1991.

\bibitem{weinberg2018topics}
G.~Weinberg, \emph{Topics in Radar Signal Processing}.\hskip 1em plus 0.5em
  minus 0.4em\relax BoD--Books on Demand, 2018.

\end{thebibliography}

\end{document}